\documentclass[conference]{IEEEtran}
\makeatletter
\def\ps@headings{%
\def\@oddhead{\mbox{}\scriptsize\rightmark \hfil \thepage}%
\def\@evenhead{\scriptsize\thepage \hfil \leftmark\mbox{}}%
\def\@oddfoot{}%
\def\@evenfoot{}}
\makeatother
\IEEEoverridecommandlockouts
\usepackage{cite}
\usepackage{amsmath,amssymb,amsfonts}
\usepackage{algorithmic}
\usepackage{graphicx}
\usepackage{textcomp}
\usepackage[utf8]{inputenc}
\usepackage{graphicx}
\usepackage{textcomp}
\usepackage{indentfirst}
\usepackage{xcolor}
\usepackage{algorithmic}
\usepackage{amsmath,amssymb,amsfonts}
\usepackage[ruled,vlined]{algorithm2e}
\usepackage{graphicx}
\usepackage[export]{adjustbox}
\usepackage{xcolor}

\usepackage{comment}  
\usepackage{dblfloatfix} 
\usepackage{adjustbox}
\usepackage[caption=false]{subfig}
\begin{document}

\title{Privacy, Security, and Utility Analysis of Differentially Private CPES Data}

\author{\IEEEauthorblockN{
Md Tamjid Hossain,
Shahriar Badsha,
Haoting Shen
    }
    
\IEEEauthorblockA{University of Nevada, Reno, NV, USA}

Email: mdtamjidh@nevada.unr.edu,\{sbadsha, hshen\}@unr.edu
}

\maketitle

\begin{abstract}
Differential privacy (DP) has been widely used to protect the privacy of confidential cyber physical energy systems (CPES) data. However, applying DP without analyzing the utility, privacy, and security requirements can affect the data utility as well as help the attacker to conduct integrity attacks (e.g., False Data Injection(FDI)) leveraging the differentially private data. Existing anomaly-detection-based defense strategies against data integrity attacks in DP-based smart grids fail to minimize the attack impact while maximizing data privacy and utility. To address this challenge, it is nontrivial to apply a defensive approach during the design process. In this paper, we formulate and develop the defense strategy as a part of the design process to investigate data privacy, security, and utility in a DP-based smart grid network. We have proposed a provable relationship among the DP-parameters that enables the defender to design a fault-tolerant system against FDI attacks. To experimentally evaluate and prove the effectiveness of our proposed design approach, we have simulated the FDI attack in a DP-based grid. The evaluation indicates that the attack impact can be minimized if the designer calibrates the privacy level according to the proposed correlation of the DP-parameters to design the grid network. Moreover, we analyze the feasibility of the DP mechanism and QoS of the smart grid network in an adversarial setting. Our analysis suggests that the DP mechanism is feasible over existing privacy-preserving mechanisms in the smart grid domain. Also, the QoS of the differentially private grid applications is found satisfactory in adversarial presence.
\end{abstract}


\section{Introduction}
\label{Intro}

\noindent Cyber Physical Energy System (CPES) such as smart grid data is used for various mission-critical (e.g., state estimation, microgrid islanding, and synchronization, load balancing, etc.) and non-mission-critical applications (e.g., energy consumption prediction, power outage forecasting, etc.) \cite{sadeghi2015security, sun2016cyber, berriel2017monthly, liu2020study}. The granular level grid data with numerous features pave the way for state-of-the-art security and privacy research. However, grid data also carries the personal and confidential information of the customers. In most cases, releasing such data is restricted due to security and legal issues \cite{fioretto2019differential}. Therefore, the data needs privacy preservation, especially during data sharing or data exchanging operations \cite{giraldo2017security}. 

Over the years, several research works have been conducted on the data privacy-preservation mechanisms (e.g., secure multi-party communication, k-anonymization,  l-diversity, etc.) \cite{kelarev2019multistage} and the privacy-violating attacks on those mechanisms \cite{narayanan2008robust,sweeney2000simple, ohm2009broken}.
To overcome such privacy-violating attacks while preserving data confidentiality and maintaining a level of data utility, the concept of DP (Differential Privacy) has been introduced by Dwork et al. \cite{dwork2006calibrating}. The DP mechanism adds sufficient randomized noise to the aggregation result. This randomized noise prevents the attacker from revealing the identity of an entity by obscuring the contribution of a single record. 
This privacy assurance can motivate the data providers to provide open access to their differentially private databases to the research community and others.
\subsection{Motivations}
\label{motive}
\noindent Although the DP mechanism achieves a privacy guarantee, a well-known drawback of it is the degradation of data utility with the increments of the data privacy \cite{dwork2006calibrating}. Moreover, from recent research by Giraldo et al., we came to know that 
differentially private data (i.e., DP-data) can be exploited by the attacker to conduct False Data Injection (FDI) attacks on a DP-based system in smart grid domain \cite{giraldo2020adversarial}. These data integrity attacks on differentially private data have led us to the question- ``\textit{What should be the design approach to defend a DP-based cyber physical energy systems (CPES) against FDI attack?}". To answer these questions, it is essential to find out the factors that facilitate this FDI attack. Likewise, to design a fault-tolerant, resilient, privacy-preserved, and well-secured defense mechanism, a provable relationship among these factors, outcomes, and parameters of the DP method needs to be developed.

Besides, any active or passive attack degrades the overall QoS of a system. So, it is also essential to raise and subsequently answer the question- ``\textit{What are the impacts of the FDI attacks on the QoS of a DP-based system?}". Based on this impact analysis of the FDI attack, the feasibility analysis of the DP technique in the CPES domain needs to be carried out in an adversarial setting. Successively, the feasibility analysis would help the CPES designers to determine the proper applicable areas of the DP mechanism.

As DP can be a tool for the attacker to conduct the FDI attack, it should not be applied in the mission-critical operations of any CPES just to enhance data privacy. On the contrary, the method must be applied with proper security measures. Otherwise, the entire CPES can be comprised by any malicious actor. Our motivation behind this research comes from these defense perspectives. 

Earlier researches have focused on building robust anomaly detection algorithms to either defend a DP-based CPES or improve the accuracy of ML-based prediction algorithm \cite{giraldo2020adversarial, lecuyer2019certified, cohen2019certified}. 
However, if the DP technique is applied in several layers of the CPES network, then identifying the false data is very difficult (if not impossible) for the traditional anomaly detectors. Therefore, novel defense strategies need to be devised against FDI attacks in DP-based smart grid applications, and for that more research should be carried out to analyze the correlation among DP-parameters. 
\subsection{Contributions}
\label{contrib}
\noindent In this paper, we address and analyze the concern of the FDI attack exploiting the differentially private data through adversarial analysis of the DP-based mechanisms into the smart grid domain.  Our main contributions in this research work can be summarized as follows:
\begin{itemize}
    \item We have demonstrated the formulation of a provable relationship between the attack impact and the parameters (data sensitivity, privacy loss, and attackers' tolerance to be detected) associated with DP mechanism. This relationship enables the designer to design a robust and fault-tolerant DP-based privacy-preserving system.
    \item We have experimentally evaluated as well as proved that the maximum attack impact of the FDI attack could be minimized if the designer uses our proposed approach to design the process in the first place.
    \item We have rigorously analyzed the feasibility of the DP mechanism in CPES data under an adversarial setting from a QoS point of view. We further evaluated the usability of the data considering an attack scenario where the attacker manipulates the data despite all the privacy and security measures. The evaluation shows that the manipulated data are usable for many non-mission-critical operations of CPESs like smart grids.  
\end{itemize}
Section \ref{LitReview} of this paper covers a brief review of the related works while section \ref{ProbState&Model} outlines the research problem and threat model. Section \ref{OptAttackOptDefense}
points out the objectives of the attacker and the defender as well as develops correlation among DP-parameters as a part of a defense against optimal FDI attack. Section \ref{ExpAnalysis} analyzes the effectiveness of our proposed model and correlation through QoS and feasibility analysis. Finally, in section \ref{ConclusionFuture}, we conclude the paper with some future research directions.

\section{Literature Review}
\label{LitReview}
\noindent To protect data confidentiality, numerous privacy-preserving techniques have been suggested and applied in the smart grid domain over the last decades \cite{liu2018practical, ferrag2016survey, lu2012eppa}. Some of them use cryptographic encryption techniques (AES, RSA, etc.), while others incorporate homomorphic encryption schemes, pseudonymization, l-diversity, t-closeness, differential privacy, etc.\cite{kelarev2019multistage}. Each of them has its advantages and shortcomings. The advantage of encryption is that it can transfer data without noises while the drawbacks are the high resource utilization, cryptographic key generation, and distribution complexity, latency, etc. \cite{lu2014toward, barbosa2016technique}. Likewise, the data anonymization techniques also suffer from several drawbacks which include the high possibility of being compromised by revealing actual identities and impracticability of the method for data containing millions of records \cite{narayanan2008robust}. 

\subsection{Countermeasures of Passive Attacks with DP technique}
\noindent Unlike other privacy-preserving techniques, DP can preserve data privacy more effectively by perturbing a small amount of noise in the query result while making sure not to overly degrade the data utility for many application scenarios. As a consequence, researchers have proposed to use the DP mechanism in CPESs like smart grids, IoT networks, automated vehicles, etc. to protect the data privacy \cite{barbosa2016technique, sandberg2015differentially}. In the smart grid domain, the DP technique has been proposed for smart meters \cite{barbosa2016technique} and state estimations \cite{sandberg2015differentially}, power grid obfuscation \cite{fioretto2019differential}, PMUs (Phasor Measurement Units) and $\mu$PMUs \cite{pinte2015low}. 
All of these works have proposed DP as a solution to passive attacks (e.g., eavesdropping). However, none of these works have considered the impact of the FDI attacks in their smart grid schemes.

\subsection{Countermeasures of Active Attacks in Smart Grid}
\label{ReviewCounter}
\noindent Another related line of works focuses on the data integrity attacks (e.g., FDI attack) on state estimation, future consumption prediction, billing, and pricing, etc. in the smart grid domain \cite{liu2011false}. Several techniques have been proposed over the years to prevent data integrity attacks using methods such as bad data filtration, blockchain, encryption, etc. \cite{rahman2013false}. In particular, the integration of blockchain to achieve security and integrity has been found effective in the smart grid sector \cite{bhattacharjee2020block, bhattacharjee2020blockchain, hossain2020porch}. 
Although these works have considered the active attacks in their schemes, they have not formulated these attacks in any DP-based system.
\begin{table*}[b!]
\centering
\caption{List of symbols and their description}
\label{table:1}
\begin{tabular}{||c l c l||} 
 \hline
 Symbols & Description & Symbols & Description \\ [0.5ex] 
 \hline\hline
 $X$ & Dataset contains all the measurement value & $\mu_a^*$ & Optimal attack impact \\ 
 $X^\prime$ & Dataset differs by a single record from $X$ & $\tau$ & Query result deviation threshold \\
 $t$ & Transcripts & $V$ & Measured voltage by PMU \\
 $\varepsilon$ & Privacy loss & $f_0$ & PDF of Laplace distribution \\
 $\Delta f$ & Data sensitivity & $f_a$ & PDF of attack distribution \\
 $Q$ & Differentially private data & $f_a^*$ & PDF of optimal attack distribution \\
 $\eta$ & Noise drawn from Laplace distribution & $\theta$ & Mean or location parameter of Laplace distribution \\
 $\eta_a$ & Noise drawn from Optimal attack distribution & $k_1$ & Kullback-Leibler divergence \\ 
 $b$ & Scale parameter of Laplace distribution & $y$ & Query result \\ 
 $\mu_a$ & Attack impact & $\gamma$ & Attackers' tolerance to be detected \\[1ex]
 \hline
\end{tabular}
\end{table*}
\subsection{Adversarial Classification of the DP technique}
\label{ReviewAdvClass}
\noindent Several studies \cite{giraldo2017security_2, farokhi2018security} have considered the active attacks (particularly, FDI attack) in a DP-based CPES (e.g., smart grid, transportation systems, etc.). In \cite{giraldo2017security_2}, Giraldo et al. have analyzed how an attacker can take advantage of the added noise in DP to design stealthy attacks that maximize the physical impact in the system. Following their research, \cite{farokhi2018security} shows that the level of guaranteed privacy times the level of security equals or upper bounded by a constant. Nonetheless, these works have mostly considered the defense mechanism based on anomaly detection schemes.

The closest work to our own is \cite{giraldo2020adversarial}, which has proposed the optimal FDI attacks that degrade the anomaly detection capabilities of the system while allowing the attacker to remain undetected by ``hiding" false data into the DP noise. The authors also proposed a detection-based defense mechanism to minimize the impact of such attacks. Contrarily to the post-attack detection, we have developed the correlation that facilitates the defense against FDI attacks. Our proposed scheme will assist the system designer to design a robust DP-based architecture considering adversarial presence.

In addition to work on the defense mechanism as a part of the design process, our work is related to the feasibility and QoS analysis of DP-based systems in an adversarial setting. To date, little attention has been paid to this line of research due to the provable privacy guarantee of the DP mechanism. In \cite{pokhrel2020qos}, a QoS-aware personalized privacy protection model has been proposed. However, their work considered privacy attacks only. \cite{jeon2011qos} has outlined the QoS requirements of smart grid in terms of low latency, seamless data availability, high data utility, and low complexity. 

Nevertheless, their architecture does not cover the DP technique and the active attack scenarios. As DP can be a tool for the attacker to conduct active attacks in the smart grid domain, it is necessary to analyze the feasibility of applying the DP mechanism in smart grid operations. At the same time, it is also essential to carry out the QoS analysis of a DP-based smart grid in the adversarial setting to determine the most suitable applications where DP can be used as a secured privacy-preserving tool. This work also focuses on this particular issue and carries out the required QoS analysis under an active attack scenario.

\section{Problem Statement and Threat Model}
\label{ProbState&Model}
\noindent In this section, the main research problem, as well as the objectives of this research, are formulated.
Also, a threat model of the research problem is developed. Table \ref{table:1} provides the symbols and their description used in this paper.
\subsection{Problem Statement}
\label{ProbState}
\noindent Fig. \ref{DPunderAttack} depicts the basic operational principle of a DP-based system. A mechanism is $\varepsilon$- indistinguishable if for all pairs $X,X^\prime \in\; D^n$ which differ in only one entry, for all adversaries $A$, and for all transcripts $t$ \cite{dwork2006calibrating}:
\begin{gather}
\ln{\left|\frac{Pr[T_A(X) = t]}{Pr[T_A(X^\prime) = t]} \right|} \leq \varepsilon
\label{dworkEqn}
\end{gather}
Equation (\ref{dworkEqn}) provides the privacy guarantee of DP mechanism.
Applying DP during a sum query over a database containing $\{x_1, x_2, x_3,..,x_i,..\}$ as input, we get-
\begin{gather}
    \sum_{i} x_i + Y; Y \sim Lap(\frac{\Delta f}{\varepsilon})
\end{gather}
Here, $Lap$  represents the Laplacian distribution mechanism. $\Delta f$ is the sensitivity of the data which is inherent in the dataset. 
For desired privacy, the noise is calibrated according to the sensitivity of the data. Larger noise yields a smaller value of privacy loss and vice versa. However, more noise leads to less data utility. Besides, applying large noise into the data for ensuring better privacy provides an attacker the opportunity of injecting false data into it. So, if the DP technique is applied in several layers of a CPES network and an attacker injects a low amount of false data into the DP-data (differentially private data), then the traditional anomaly detectors cannot detect the manipulation.

An alternative defense strategy that can minimize the attack impact is to design and apply the DP technique targeting minimum attack impact and maximum privacy.
For this, a verifiable correlation between the outcome (attack impact) and the parameters of the DP method needs to be developed that can enable the designer to design a resilient and privacy-preserved system as a part of the defense mechanism.

The primary objective of our analysis is to enable the defender to design a privacy-preserved and secured DP-based system (e.g., synchrophasor network) against FDI attacks. The secondary objective of our research is to analyze the feasibility of the DP mechanism in synchrophasor networks considering the vulnerability of the DP-based system against FDI attacks. We also analyze the QoS of the DP-based synchrophasor network in an adversarial setting. More precisely, we want to raise and subsequently answer the following questions.

\subsubsection{Design objective}
\label{DesignObj} Given a dataset $(D)$, what will be the value of scale parameter $(b)$ and epsilon $(\varepsilon)$ such that the attack impact $(\mu_a)$ becomes equal or close to the actual query result (i.e., $\mu_a = Q(D) \vee \mu_a \to Q(D)$)
\subsubsection{Feasibility analysis objective}
\label{FeasibilityObj} Would the DP technique remain feasible over other privacy-preserving mechanisms (e.g., encryption, masking, anonymization, etc.) considering its security challenges (i.e., the chance of FDI attacks exploiting DP-noise)?
\subsubsection{QoS analysis objective}
\label{QoSObj} What would be the QoS of the DP-based synchrophasor network in terms of computational overhead and data utility under the FDI attack?


\subsection{Threat Model}
\label{ThreatModel}
\begin{figure}[!b]
    \centerline{\includegraphics[width=\linewidth]{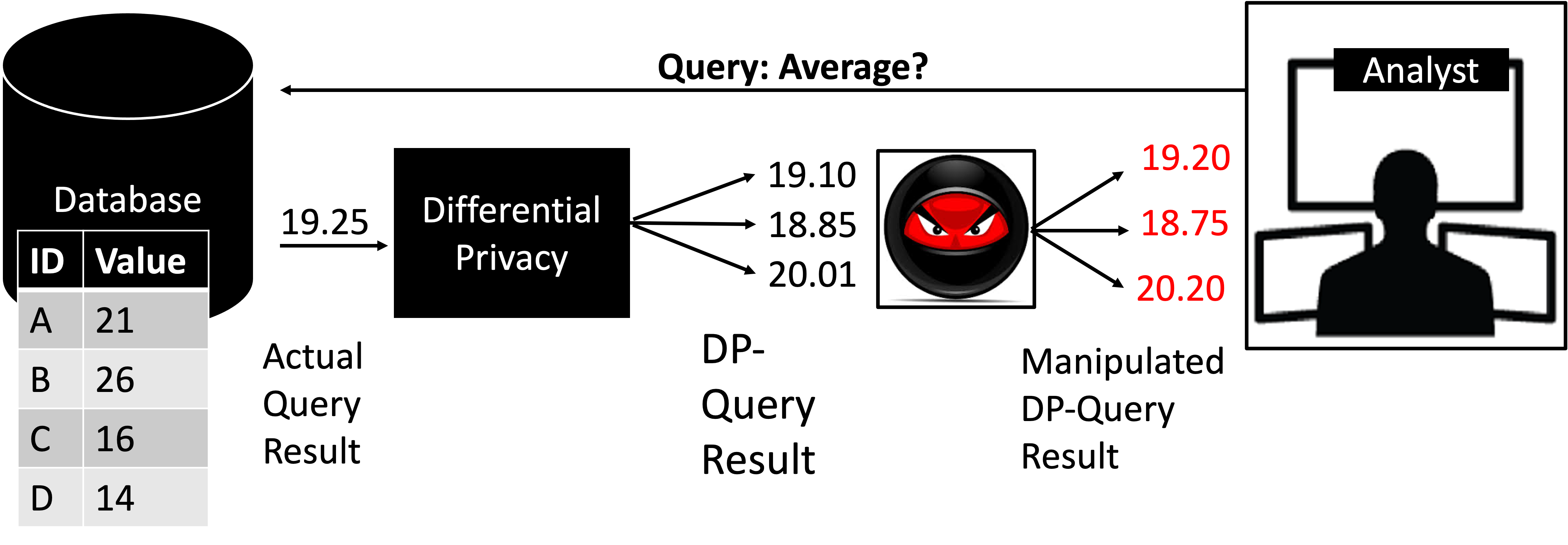}}
    \caption{Differential privacy in an adversarial setting. Differentially private query results are manipulated by a malicious actor and sent to the analyst. Small amount of manipulation is difficult to detect}
    \label{DPunderAttack}
\end{figure}

\noindent The proposed threat model, as depicted in Fig. \ref{DPunderAttack}, considers the FDI attack on the data packets while transferring from one node to another node through the spoofing technique. Throughout the attack, the attacker pretends to be an honest node to all other nodes and injects false measurement as a form of noise into DP-data. The model is developed based on an attack scenario, where the attacker can conduct the FDI attack despite all the privacy and security measures. 
More formally, for a DP-result of any aggregated query over a database consisting of grid measurement, the FDI attack is expressed as follows:
\begin{gather}
Q_a = Q(D) + \eta_a\; s.t.\; Q(D) = \sum_{i=1}^{n}{v_i} + \eta\; \forall\; v_i \in V
\label{OptAttackDist}
\end{gather}
where, $V=\{v_1,v_2,...,v_n\}$ is the dataset containing $n$ number of measurement values obtained from a $\mu$PMU dataset, $\eta$ is the measure of Laplace noise, $Q(D)$ is the non-manipulated differentially private query result over database $D$, $\eta_a$ is the false measurement injected by the attacker as a form of noise and $Q_a$ is the manipulated query result. 
Moreover, the Laplace noise, $\eta$ is a random variable with a probability distribution that satisfies the condition of the differential privacy (as stated by (\ref{dworkEqn})). Other than the Gaussian and the Exponential distribution, the Laplace distribution is also a good choice for extracting random noise as it satisfies $(\varepsilon,0)$ or  $(\varepsilon,\delta)$ – differential privacy. Also, the Probability Density Function (PDF) of Laplace distribution has a fatter tail which is why Laplacian noise provides better privacy. The Laplace distribution with mean $(\theta=0)$, variance $(\sigma^2=2b^2)$ and PDF, $f_0$ can be described by (\ref{foPDF}).
\begin{gather}
f_0 = \frac{1}{2b^2}\exp{\frac{-\left|y-\theta\right|}{b}} \; s.t.\; b = \frac{\Delta f}{\varepsilon}
\label{foPDF}
\end{gather}

Here, $b$ is the scale parameter and can be represented as the ratio between the sensitivity of the data $(\Delta f)$ and the privacy loss $(\varepsilon)$.
We consider that the attacker varies the value of $\eta_a$ according to her attack tolerance level (i.e., the willingness to be detected or remain undetected). Now, if $\eta_a$ is sufficiently small, then it is unlikely for the recipient to detect the manipulation without a well-designed and robust anomaly detector. On the other hand, the anomaly detectors also add extra latency to the network and slow down the system performance. This is a potential threat to the system's performance that needs to be minimized. 

\section{Optimal Attack and Defense Strategy}
\label{OptAttackOptDefense}
\noindent In this section, we first describe the attackers' and the defenders' objectives. We then compute the correlation among the DP-parameters to facilitate defense against the optimal attack. 
\subsection{Attackers' Objective}
\label{AttackersObj}
\noindent An attacker can be a passive attacker or an active attacker. A passive attacker eavesdrops on the communication path while an active attacker directly interferes with the data (e.g., masquerades, modification of messages, denial of services, etc.). The DP mechanism gives protection against passive attacks.  However, the attacker can still modify the DP-data by injecting false data into them, which becomes a concerning security issue.

Generally, the attacker of our threat model has two major objectives while carrying out the attack– ``\textit{(a) maximum damage (b) avoid detection}". These two goals are contradictory with each other in the sense that it becomes difficult for an attacker to hide the attack or hide her identity if she wants to carry out maximum devastation to the system.
\subsection{Defenders' Objective}
\label{DefendersObj}
\noindent Defenders' objective is to defend the system against passive and active attacks. The defense can be carried out from various fronts of the system. One such technique can be setting up the anomaly detectors to minimize the attack impact. But, this would add extra latency to the system performance. Another way could be designing the system as a specific fault-tolerant in the first place considering the possible FDI attacks in the future. In our model, the defender has two major objectives- ``\textit{(a) design the process as fault-tolerant against FDI attacks (b) preserve the data privacy and QoS of the system}".

These specific design requirements have led us to the quest of finding a provable relationship among the DP-parameters to achieve desired privacy, utility, security as well as the feasibility of the DP mechanism.

\subsection{Optimal Attack Strategy}
\label{OptAttack}
\noindent An optimal attack would take place when the objectives of the attacker are achieved with maximum possible payoffs. In \cite{giraldo2020adversarial}, Giraldo et al. have discussed this optimization problem. They shows that an optimistic attacker would draw noise $\eta_a$ from a probability density function, $f_a^*$, which can be represented as follows:
\begin{gather}
f_a^* = \frac{k_1^2 - b^2}{2bk_1^2}\exp{\frac{-\left|y-\theta\right|}{b} + \frac{(y-\theta)}{k_1}}
\label{OptAttackEq}
\end{gather}
\indent Here, $b$ is the scale factor and $\theta$ is the location parameter. $k_1$ is a solution to KL (Kullback–Leibler) divergence between two probability distributions $D_{KL}(f_a\left|\right|f_0)$ and can be found by solving below equation –
\begin{gather}
\frac{2b^2}{k_1^2 - b^2} + ln(1 - \frac{b^2}{k_1^2}) = \gamma \;\;\; \forall \; k_1 > b
\label{AttackK1}
\end{gather}
\indent Here, $\gamma$ is the attackers’ tolerance to be detected. A large $\gamma$ would mean that the attacker does not care to be detected whereas a small $\gamma$ means hiding her entity is crucial for the attacker. To achieve her second objective which is to remain undetected as long as possible the value of $\gamma$ should be as small as zero. However, then the attackers’ probability distribution would be similar to the original probability distribution of the Laplace mechanism (i.e., $f_a^* \simeq f_0$). So, the attacker must go for a trade-off between the two objectives. If the attacker chooses a small value of $\gamma$, i.e., to remain undetected as to fulfill the second objective, $k_1$ approaches infinity. Again, as $k_1$ approaches to infinity, the PDF, $f_a^*$ as described by (\ref{OptAttackEq}) approaches to $f_0$. The optimal attack impact is given by the following formula \cite{giraldo2020adversarial}.
\begin{gather}
\mu_a^* = \frac{b^2 (\theta - 2k_1) -\theta k_1^2}{b^2 - k_1^2}
\label{AttackMu}
\end{gather}
\indent Here, $\mu_a^*$ is the amount of optimal bias introduced by the attacker. It depends on four parameters (mean, $\theta$; attacker’s tolerance, $\gamma$; data sensitivity, $\Delta f$ and privacy loss, $\varepsilon$). The optimal attack impacts $(\mu_a^*)$ largely vary with the privacy loss and attackers’ tolerance. More specifically, a low privacy loss (or a large amount of noise) and a high attackers’ tolerance help the attacker to conduct a more devastating attack (i.e., the high value of $\mu_a^*$) by injecting large false data and hiding behind the large noise of differential privacy.

\subsection{Parameter Design for Effective Defense Mechanism}
\label{ParaDesign}
\noindent Among the four parameters, the mean ($\theta$, also called the ‘location’ in PDF) and the data sensitivity ($\Delta f$) can be calculated from the targeted dataset. Hence, these two parameters are not adjustable by the user as per the design requirements. However, these parameters are crucial for tuning the noise while applying differential privacy so that, either, the attack impact cannot be large, or the attack can be easily detected.

During the design process, the attackers’ tolerance ($\gamma$) should be selected in such a way that considers the maximum possible attack impact or deviation from the actual data. The possible incidents that can take place are given below.
\begin{itemize}
    \item The attacker chooses a very large $\gamma$ to maximize the attack impact without paying much attention to detection. In that case, a carefully designed anomaly detector can easily identify the attack. 
    \item The attacker chooses a very small $\gamma$ to minimize the disclosure of her identity.
    In that case, the attack impact or deviation would be negligible (as $f_a^*$ becomes equal to $f_0 $) and can be ignored.
    \item The attacker chooses an optimal value of $\gamma$ for an optimal attack impact and identity disclosure.
    In that case, the attack impact is not only tough to detect (if not impossible) but also impractical to be neglected totally. 
    We elaborate and analyze this point in section \ref{ExpAnalysis}.
\end{itemize}
We are now ready to present one of the main contributions (i.e., correlation among the DP-parameters considering adversarial presence) of the paper.

\textit{Theorem 4.1}: For any optimal attack strategy (stealthiness: $\gamma$, attack impact: $\mu_a^*$), the correlation among the DP-parameters that solves the design problem stated by \ref{DesignObj} can be represented as follows:
\begin{gather}
  \varepsilon = \frac{\Delta f}{b} \; s.t.\; b = \sqrt{\frac{k_1^2(\mu_a^* - \theta)}{2k_1 + (\mu_a^* - \theta)}} \;\; \forall \left| \mu_a^* - \theta \right| \geq 0
    \label{theorem}
\end{gather}

\textit{Proof}: The scale factor, $b$ needs to be greater than zero, otherwise (\ref{foPDF}) becomes undefined. Moreover, according to the optimal attack strategy described in \ref{OptAttack}, the optimal attack follows the Laplace distribution. Therefore, the absolute deviation of the attack impact from the mean of the query data (alternatively, Laplace density) cannot be less than zero (i.e., $\left| \mu_a - \theta \right| \geq 0)$). Under these conditions, we solve for the scale parameter, $b$ from (\ref{AttackMu}) following simple algebraic rules and get–
\begin{gather}
    b = \sqrt{\frac{k_1^2(\mu_a^* - \theta)}{2k_1 + (\mu_a^* - \theta)}} \;\;\; \forall \left| \mu_a^* - \theta \right| \geq 0
    \label{Designb}
\end{gather}
Since it is an optimization problem in a constrained environment, we model $k_1$ as a Lagrange multiplier and compute $k_1$ numerically using (\ref{AttackK1}) and (\ref{Designb}). Thus, the general expression of (\ref{AttackK1}) can be rewritten as (\ref{Designk}). Moreover, we can solve $k_1$ numerically from (\ref{Designk}) and successively compute $\varepsilon$ from (\ref{DesignEps}).

\begin{gather}
\label{Designk}
    \frac{\mu_a^* - \theta}{k_1} + \ln{\left|\frac{2k_1}{2k_1 + (\mu_a^* - \theta)}\right|} = \gamma\\
    \varepsilon = \frac{\Delta f}{b}
    \label{DesignEps}
\end{gather}



We can further evaluate the proposed theorem through boundary conditions.

\textit{Case-1 (lower bound, $\mu_a^* - \theta \to 0$)}:
When $\mu_a^* - \theta \to 0$ (i.e., when we want to minimize the attack impact and make the deviation zero), (\ref{Designk}) becomes-
\begin{align}
    &0 + \ln{\frac{2k_1}{2k_1 + 0}} = \gamma \implies \gamma \to 0
\end{align}
Here, $\gamma \to 0$ indicates that the attacker does not want to be detected at all and hence the payoff of the attacker (i.e., $\mu_a^* - \theta$) approaches to zero. After rewriting (\ref{Designb}), we can apply limit ($\mu_a^* - \theta$)$\to0$ as follows:

\begin{align}
\lim_{(\mu_a^* - \theta)\to0}{b} = \lim_{(\mu_a^* - \theta)\to0}{\sqrt{\frac{k_1^2}{\frac{2k_1}{\mu_a^* - \theta} + 1}}} \implies b \to0
\label{DesignEpsInf}
\end{align}

\textit{Remark 4.1}: When $b\to0$, then $\varepsilon \to\infty$ since $\varepsilon = \frac{\Delta f}{b}$
Equation (\ref{DesignEpsInf}) and Remark 4.1 indicates that if the privacy loss, $\varepsilon$ approaches to infinity (alternatively, noise approaches to zero) and the attacker does not want to be detected ($\gamma \to 0$), then the attack impact is minimum and the deviation from the mean is close to zero (i.e., $\mu_a^* - \theta \to 0$).

\textit{Case-2 (upper bound, $\mu_a^* - \theta \to\infty$)}:
If we consider the maximum attack impact (i.e., $\mu_a^* - \theta \to\infty$), from (\ref{Designk}), it can be inferred that $\gamma \to\infty$, which means the attacker does not care about to be exposed and goes for maximum attack impact. From \ref{OptAttack}, we know that, when $\gamma \to\infty$, then $k_1\to0$. Now, when $k_1\to0$ is very small and $(\mu_a^* - \theta) \to \infty$, then the scale factor ($b$) is very large. 

\textit{Remark 4.2}: When $b\to\infty$, then $\varepsilon \to0$ since $\varepsilon = \frac{\Delta f}{b}$.
Therefore, if the privacy loss, $\varepsilon$ approaches to zero (0) (alternatively, noise approaches to infinity) and the attacker does not care to be exposed ($\gamma \to \infty$), then the attack impact is maximum and the deviation from the mean very large (i.e., $\mu_a^* - \theta \to \infty$).

In summary, (\ref{Designb}) – (\ref{DesignEps}) can be used to adjust the privacy loss ($\varepsilon$) for a given sensitivity, $\Delta f$, an optimal attack impact, $\mu_a^*$, and a level of attackers’ tolerance, $\gamma$. The defender can use the correlation of (\ref{Designb}) and (\ref{DesignEps}) to design a robust, fault-tolerant, privacy-preserved, and well-secured DP-based system. We discuss the effectiveness of our proposed design approach analytically in section \ref{ExpAnalysis}.

\subsection{Modelling the Criterion for Feasibility and QoS Analysis}
\label{ModellingQoSFes}
\noindent Analyzing QoS (Quality of Service) of the DP-based system under attack scenario is another key objective of our research. Here, QoS refers to the measurement of the overall performance of the system. To quantitatively measure the QoS of the system, several parameters of the network are considered, such as overall latency, congestion, data availability, data utility, system complexity, network speed, operational overhead, data redundancy, system resiliency, etc. As we have not considered any data availability attacks (e.g., DDoS attack, delay attack, etc.) in our model, we have eliminated the data availability from our QoS criterion list. Moreover, the DP technique is a better choice than other privacy-preserving techniques in terms of system complexity, system resiliency, and operational overhead. The cryptographic encryption techniques incorporate key generation, key distribution, and third-party involvement while the anonymization techniques require huge time for a large dataset. Consequently, we measure the QoS of the attack-prone DP-based synchrophasor network through overall data utility along with the privacy and security cost of DP.

As for the feasibility analysis, it is an assessment of the practicability of a proposed plan. The analysis is carried out by asking and subsequently answering the question- ``\textit{Is the proposed method feasible?"}. We analyze this feasibility question under an adversarial scenario. A method is labeled as feasible if it preserves the specific requirements (e.g., privacy, security, etc.) as well as the QoS of the system. From this perspective, the DP technique is feasible for the synchrophasor network if it provides a substantial amount of data privacy, does not violate the security of the system, and does not degrade the QoS of the system below a satisfactory level. At the same time, it is also necessary to compare the DP technique with other existing privacy-preserving techniques in terms of computational overhead. We carry out such a comparison in section \ref{ExpAnalysis} of this paper.

\section{Experimental Analysis}
\label{ExpAnalysis}
\noindent For the experimental purpose, we have used the same dataset used by \cite{pignati2015real}. The dataset includes the real-time measurement of dedicated PMUs connected on the medium voltage side of the network secondary substations in a smart grid on the EPFL campus. We have selected this data because similar data are being used in various applications in the smart grid domain, e.g., state estimation, modeling power network, predicting future needs, etc. The data contains historical measurement values from 5 PMUs starting from the year 2014 to 2019 with a high data missing percentage. However, since the frequency of the measurement is the fraction of seconds (ms), the size of the dataset is sufficient for the practical demonstration purpose of our proposed model.  

\subsection{FDI Attack Simulation in a Synchrophasor Network}
\label{FDISimulation}
\noindent The FDI attack can take place in different nodes (master nodes, fog nodes, edge nodes, etc.) of a synchrophasor network. The attacker can conduct the attack either by compromising nodes (or sensors) through direct access or by compromising the communication channel through the traditional MITM (Man-in-the-Middle) attack approach \cite{yang2012man}. As physical access to a grid network is difficult for an attacker to achieve, an FDI attack by physically accessing the node is unlikely to occur. Therefore, for our proposed model we have considered the FDI attacks conducted by compromising the communication channel of the synchrophasor network. 

In a synchrophasor network, the queries among the master nodes (controller), fog nodes (PDCs-Phasor Data Concentrates), and edge nodes (PMUs) flow from the master nodes towards the edge nodes. 
\begin{figure}[!b]
    \centerline{\includegraphics[width=\linewidth]{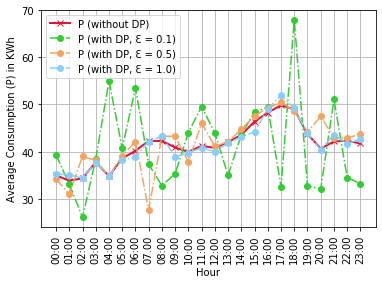}}
    \caption{Hourly average user consumption (KWh) recorded by a PMU. Small level of privacy loss ($\varepsilon$) leads to higher deviation. The DP-noise has been drawn from a Laplace distribution with zero mean and $2b^2$ variance $(\theta = 0, \sigma^2=2b^2)$}
    \label{AvgPower}
\end{figure}
\begin{figure*}[b!]
    \centerline{\includegraphics[width=\linewidth]{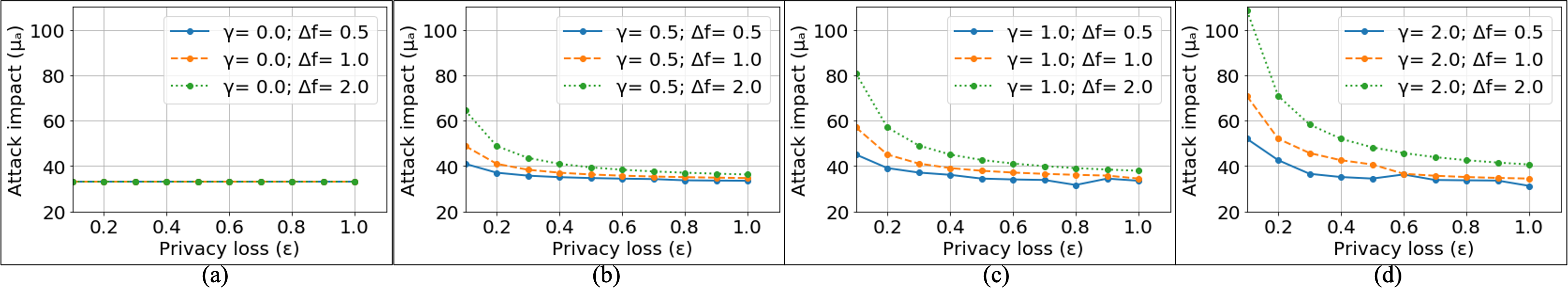}}
    \caption{Correlation between the attack impact $(\mu_a)$ and the DP-parameters. Attack impact $(\mu_a)$ decreases when- (i) privacy loss $(\varepsilon)$ increases, (ii) attackers' tolerance $(\gamma)$ decreases, and (iii) data sensitivity $(\Delta f)$ decreases.}
    \label{ImpactIncrease}
\end{figure*}
When DP is applied over the entire synchrophasor network, hierarchically, the receiving node (e.g., PDC) gets the DP-data from the lower-level nodes (e.g., PMUs). However, as there are multiple layers in the synchrophasor network, the DP technique adds random noise multiple times in the same query result. So, it becomes difficult for the anomaly detectors to identify the false data.

In our experiment, the query is made to find out the hourly average of the user consumption, P (in KWh) recorded by a PMU on a particular day. Fig. \ref{AvgPower} depicts the differentially private results (DP-results) of the query along with the actual result for varying privacy loss ($\varepsilon$). The DP-results differ from the actual query results due to the random Laplacian noise added through the DP process. For demonstrating the FDI attack, some false measurement is injected deliberately as a form of noise into the DP result. The false measurement is drawn from the attackers' probability distribution following the optimal attack strategy of section \ref{OptAttack}. Corresponding attack impact is illustrated in Fig. \ref{ImpactIncrease} for various level of $\varepsilon$. The attack impact decreases for the following cases: (i) the privacy loss increases, (ii) attacker's tolerance decreases, and (iii) data sensitivity decreases.

\subsection{Privacy, Security, and Utility Analysis}
\label{PSUAnalysis}
\noindent According to the $\varepsilon$-indistinguishability condition, a DP-result differs from the actual query result (Fig. \ref{AvgPower}). When the noise is large (alternatively, $\varepsilon$ is lower), the DP-result of average user consumption deviates significantly from the true result and vice versa. The amount of random noise also depends on the data sensitivity. Highly sensitive data requires a large amount of noise to preserve data privacy. On the contrary, low sensitive data needs a small amount of noise.

Consequently, for a small amount of noise (alternatively, a high value of $\varepsilon$), the DP-result differs from the true result only by small fractions (curve without DP and with $\varepsilon = 1.0$ in Fig. \ref{AvgPower}). So, for the non-critical applications requiring less sensitive data in the smart grid domain, the required amount of noise to preserve the data privacy is also low.

\subsubsection{Correlation between the Attack Impact and the DP-parameters}
\label{ParamRelation}
The data sensitivity level has an important role in the DP mechanism. If the defender designs to add large noise to highly sensitive data, then the attacker can conduct devastating attacks. For example, in Fig. \ref{ImpactIncrease}, the privacy is very large when $\varepsilon = 0.1$. Also, for highly sensitive data (e.g., $\Delta f = 2.0$), if the defender keeps the privacy loss low ($\varepsilon = 0.1 $) to preserve privacy, the attack impact is as large as $110$. However, the attacker may not create a large attack to avoid the attack detection; rather she would follow the optimal attack distribution described by (\ref{OptAttackEq}) for an optimal payoff.

In our experiment, the true query result (true average user consumption) is $33.18$. Following the optimal attack distribution interpreted by (\ref{OptAttackEq}), the attacker can add maximum noise of $76.82$ if her tolerance level ($\gamma$) is $2.0$. Now, if $\gamma$ is increased by $0.5$ (e.g., $0.5$ to $1.0$) while $\Delta f$ remains same (e.g., $0.5$), the maximum possible attack impact is increased by approximately $4.0$ (e.g., $41.0$ to $45.0$). On the contrary, if $\Delta f$ rises by $0.5$ (e.g., $0.5$ to $1.0$) while $\gamma$ remains same (e.g., $0.5$), the attack impact increases by approximately $8.0$ (e.g., $41.0$ to $49.0$). So, it is perceived that the attack impact depends more on $\Delta f$ than $\gamma$. 

Moreover, we see that the actual mean ($\theta$) has no significant effect on the attack impact. For any mean $(\theta)$, the attack impact $(\mu_a)$ follows the same principle. That means if the mean is lower than $33.18$, the attack impact also decreases by the same amount following the DP principle. Based on this correlation between the data sensitivity $(\Delta f)$, the attackers' tolerance $(\gamma)$, and privacy loss $(\varepsilon)$, the designer can design his system targeting minimal attack impact, high data privacy, and high data utility. More specifically, the designer can calibrate the noise following (\ref{Designb})-(\ref{DesignEps}) and adjust the tolerance of the system up to the calculated attack impact value.

\subsubsection{QoS of the DP-based Synchrophasor Network in Adversarial Setting}
\label{SysQoS}
We use the defense cost and the data utility as indicators to evaluate the QoS of the DP-based system in an adversarial setting. The defense cost is the overall cost that arises from the defensive measures against cyber-attacks. It represents the distance of the DP-data from actual data. It is consists of two separate costs- the privacy cost and the security cost. Privacy cost (i.e., the distance of the DP-data from actual data due to privacy measures) takes place when the DP technique is applied to the query result. It increases with the number of times DP mechanism has been applied in the nodes. On the other hand, the security cost is added to the defense cost when the FDI attacks occur. Besides analyzing the defense cost of a DP-based synchrophasor network under an attack framework, we also measure the data utility in terms of prediction accuracy. The data utility measurement is performed with both the DP-data and the FDI-DP-data (i.e., the DP-data that is manipulated by the FDI attack).

Motivated by \cite{yu2015towards}, we build the proposed prediction model to evaluate the above-mentioned properties. The model simply executes a time-series prediction algorithm on the PMU dataset for different levels of privacy loss ($\varepsilon$). The experiment is conducted with the original data, the DP-data, and the FDI-DP-data. After executing the experiment with DP-data, the defense cost (i.e., the privacy and the security cost) is measured on the predicted values. Next, the FDI attack is conducted deliberately on both datasets (with and without DP mechanism). After that, the manipulated datasets are fed into the prediction model. Finally, the security cost (i.e., the prediction deviation) due to the FDI attack on the DP-data is measured. 

\begin{figure}[!b]
    \centerline{\includegraphics[width=\linewidth]{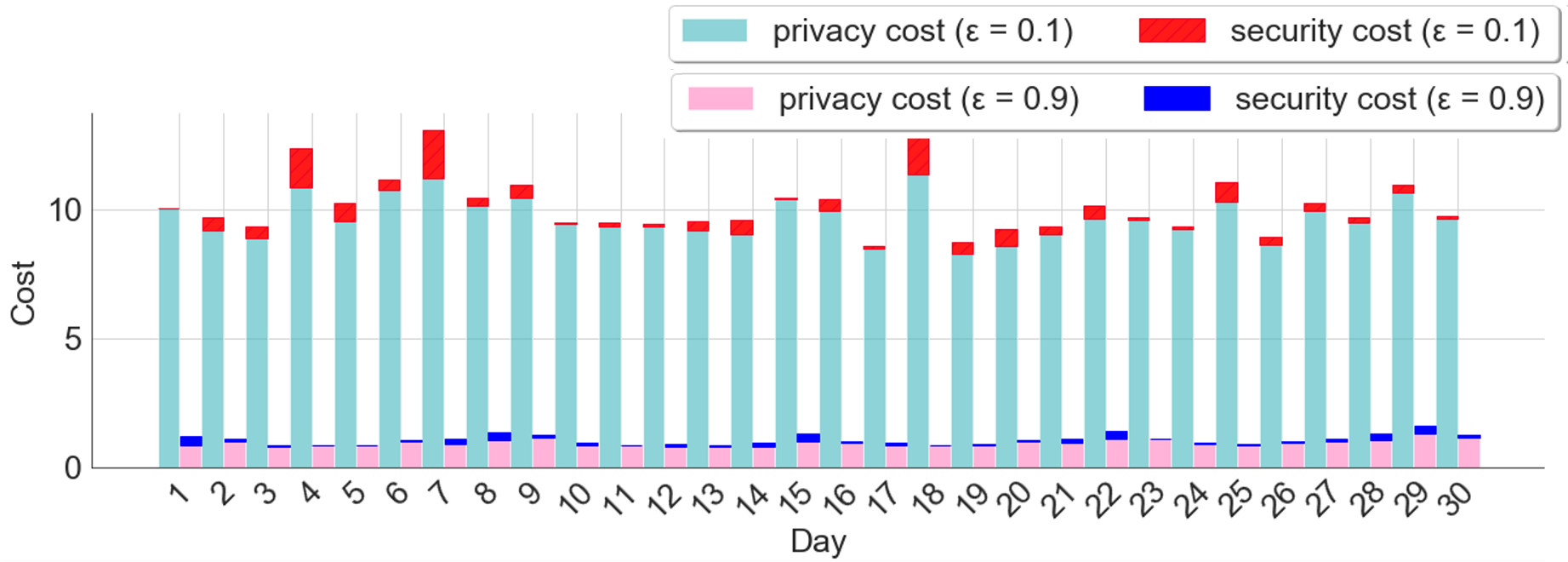}}
    \caption{Utility cost analysis of DP method for varying privacy loss ($\varepsilon$). Privacy cost is higher than the security cost. Moreover, the overall cost is small when epsilon is large (i.e., $\varepsilon = 0.9$)}
    \label{costAnalysis}
\end{figure}

\begin{figure}[!b]
\centering
 
\subfloat(a){
	\label{subfig:Original_DP}
	\includegraphics[width=\linewidth]{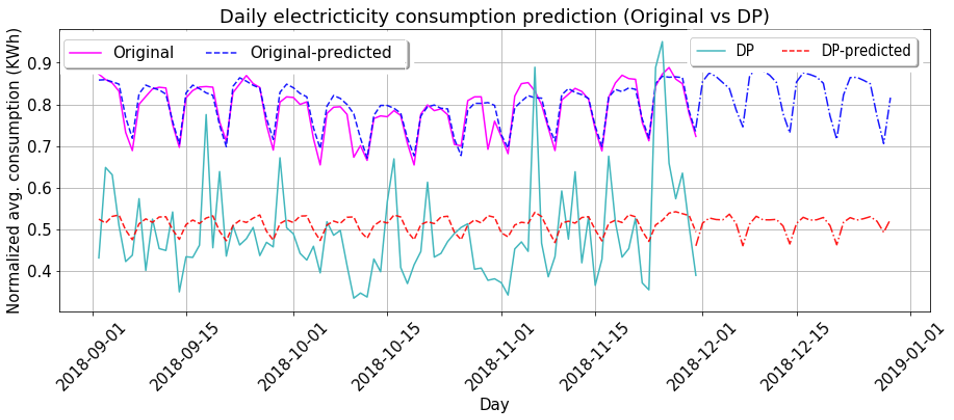} } 
\subfloat(b){
	\label{subfig:Original_FDI}
	\includegraphics[width=\linewidth]{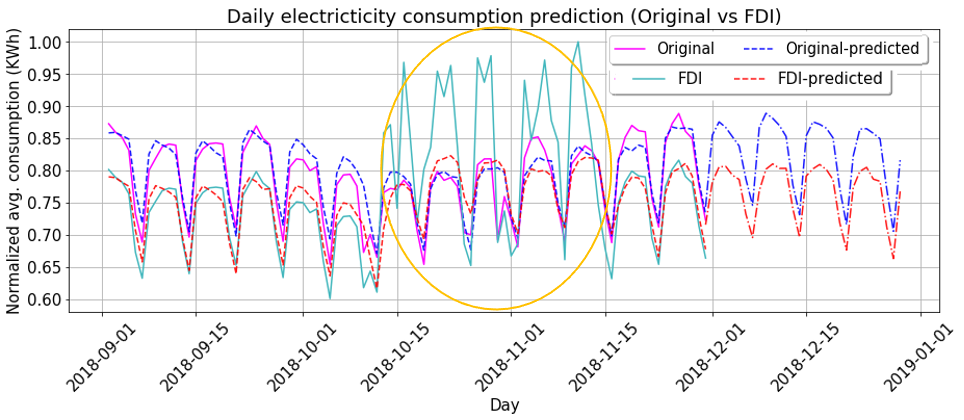} } 
\subfloat(c){
	\label{subfig:DP_FDI_DP}
	\includegraphics[width=\linewidth]{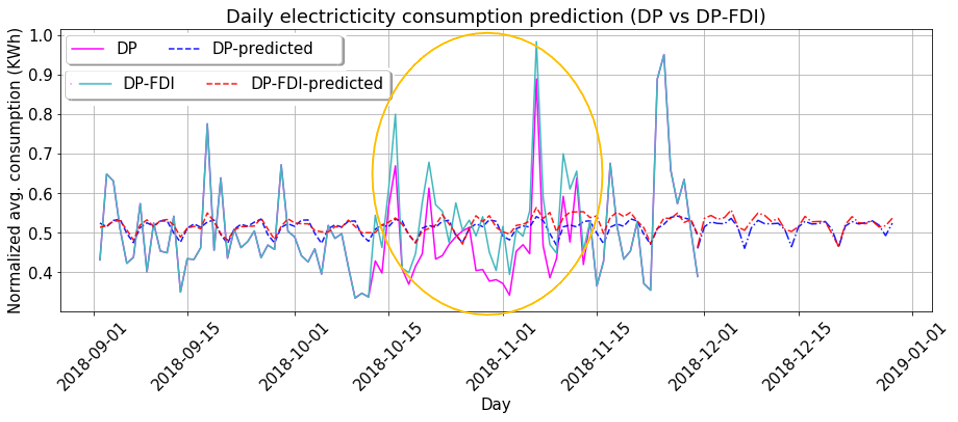}} 
\caption{QoS analysis of a DP-based system under attack scenario. (a) 'DP-predicted' vary from the 'original-predicted' due to privacy cost (b) The attack impact on the prediction accuracy is negligible for a short-time FDI attack (circled in the graph) (c) The 'DP-FDI-predicted' values are very close to the 'DP-predicted' values as the security cost is low.}
\label{QoSAnalysis}
 
\end{figure}

We first study the defense cost (i.e., the sum of privacy cost and security cost) and find that the privacy cost is higher than the security cost when the attacker conducts a stealthy attack (as shown in Fig. \ref{costAnalysis}). However, the overall defense cost reduces significantly when the privacy loss ($\varepsilon$) is increased.
Then, we investigate the utility of both the DP-data and FDI-DP-data through a time-series prediction algorithm. We find that the  ``DP: predicted" value is higher than the ``Original: predicted" value due to the privacy cost of DP mechanism as shown in Fig. \ref{QoSAnalysis}(a). After that, we deliberately conduct the FDI attack (from ``2018-10-14" to ``2018-11-14" in Fig. \ref{QoSAnalysis}(b) on the original data. Note that the FDI attack has a significant impact on the predicted values. Finally, the same degree of FDI attack is conducted intentionally on the DP-data to find the changes in the predicted values. The findings indicate that the distance between prediction accuracy with DP-data and FDI-DP-data is nominal as the security cost is very low (Fig. \ref{QoSAnalysis}(c)). We find that the impact of the FDI attack for a short period of time is low on the predicted values and can be negligible for the non-critical applications of the smart grid domain.

\subsubsection{Feasibility of DP mechanism in Synchrophasor Network}
\label{DPFeasibility}
The criterion for our feasibility analysis of DP mechanism is the comparison of overall latency. The latency comparison is carried out between the DP technique and the AES-256 encryption technique under an attack scenario. AES-256 is chosen because it is computationally faster than most of the other encryption techniques (e.g. RSA). Also, the anonymization techniques incur more latency than AES-256 encryption over the large dataset. So, for the feasibility analysis of the DP technique under attack scenario, the latency comparison between the DP and AES-256 technique should suffice. 

Both the passive and the active attacks add a latency burden to the overall system performance. In the case of the FDI attack, the attacker manipulates the data by adding false information. This process takes some time and adds latency to data flow speed. For any encryption technique, to avoid detection, the attacker needs to decrypt the data, inject false data and again encrypt the falsified data with the cryptographic key. However, for DP technique, the attacker only injects false data and remains undetected. As a consequence, the DP mechanism does not add as much latency as AES-256. Our experimental evaluation also supports this latency comparison under the FDI attack scenario. We find that the DP technique is approximately $121.49$ times faster (on an average) than the AES-256 in the same adversarial setting (i.e., the computational time of DP is only $0.0016$ seconds whereas the computational time of AES-256 is $0.1944$).

\section{Conclusion and Future Work}
\label{ConclusionFuture}
\noindent The DP mechanism can be exploited as a tool for conducting the FDI attack. To overcome this challenge, numerous research have been carried out on developing a defense mechanism based on the anomaly detection schemes in a DP-based CPES. However, the traditional anomaly detectors cannot detect the bad data added by exploiting the DP mechanism in multiple layers of the synchrophasor network. Addressing this, in this paper, we formulate and develop the defense strategy as a part of the design process. We have proposed a provable correlation among the factors affecting DP mechanism which enables the defender to design his DP-based system fault-tolerant against the FDI attacks. To experimentally evaluate and prove the effectiveness of the proposed correlation, we have simulated the FDI attack in a DP-based synchrophasor network. The evaluation indicates that the attack can be minimized using the proposed correlation in the design process. The QoS analysis of the DP-based CPES indicates the applicability of the DP mechanism in many non-critical operations. Furthermore, the feasibility analysis of DP mechanism in the CPES domain infers that the DP technique is feasible over other privacy-preserving mechanisms in terms of computational overheads. 

DP has the potential to preserve data privacy in other fields also such as edge computing and cloud computing \cite{du2018differential, wang2020edge}.  We hope that this research will contribute to those fields too. Continuing this line of research, we hope to analyze the impact of distributed active attacks on the DP-based system. In particular, we want to investigate the impact of distributed FDI attacks on personalized-privacy-aware grid architecture. Additionally, we hope to extend our model from a game-theoretic viewpoint where multiple defenders and attackers play dynamic strategies in a repeated manner. 

\bibliographystyle{IEEEtran}
\bibliography{bibliography.bib}

\begin{thebibliography}{10}
\providecommand{\url}[1]{#1}
\csname url@samestyle\endcsname
\providecommand{\newblock}{\relax}
\providecommand{\bibinfo}[2]{#2}
\providecommand{\BIBentrySTDinterwordspacing}{\spaceskip=0pt\relax}
\providecommand{\BIBentryALTinterwordstretchfactor}{4}
\providecommand{\BIBentryALTinterwordspacing}{\spaceskip=\fontdimen2\font plus
\BIBentryALTinterwordstretchfactor\fontdimen3\font minus
  \fontdimen4\font\relax}
\providecommand{\BIBforeignlanguage}[2]{{%
\expandafter\ifx\csname l@#1\endcsname\relax
\typeout{** WARNING: IEEEtran.bst: No hyphenation pattern has been}%
\typeout{** loaded for the language `#1'. Using the pattern for}%
\typeout{** the default language instead.}%
\else
\language=\csname l@#1\endcsname
\fi
#2}}
\providecommand{\BIBdecl}{\relax}
\BIBdecl

\bibitem{sadeghi2015security}
A.-R. Sadeghi, C.~Wachsmann, and M.~Waidner, ``Security and privacy challenges
  in industrial internet of things,'' in \emph{2015 52nd ACM/EDAC/IEEE Design
  Automation Conference (DAC)}.\hskip 1em plus 0.5em minus 0.4em\relax IEEE,
  2015, pp. 1--6.

\bibitem{sun2016cyber}
C.-C. Sun, C.-C. Liu, and J.~Xie, ``Cyber-physical system security of a power
  grid: State-of-the-art,'' \emph{Electronics}, vol.~5, no.~3, p.~40, 2016.

\bibitem{berriel2017monthly}
R.~F. Berriel, A.~T. Lopes, A.~Rodrigues, F.~M. Varejao, and
  T.~Oliveira-Santos, ``Monthly energy consumption forecast: A deep learning
  approach,'' in \emph{2017 International Joint Conference on Neural Networks
  (IJCNN)}.\hskip 1em plus 0.5em minus 0.4em\relax IEEE, 2017, pp. 4283--4290.

\bibitem{liu2020study}
T.~Liu, Z.~Tan, C.~Xu, H.~Chen, and Z.~Li, ``Study on deep reinforcement
  learning techniques for building energy consumption forecasting,''
  \emph{Energy and Buildings}, vol. 208, p. 109675, 2020.

\bibitem{fioretto2019differential}
F.~Fioretto, T.~W. Mak, and P.~Van~Hentenryck, ``Differential privacy for power
  grid obfuscation,'' \emph{IEEE Transactions on Smart Grid}, vol.~11, no.~2,
  pp. 1356--1366, 2019.

\bibitem{giraldo2017security}
J.~Giraldo, E.~Sarkar, A.~A. Cardenas, M.~Maniatakos, and M.~Kantarcioglu,
  ``Security and privacy in cyber-physical systems: A survey of surveys,''
  \emph{IEEE Design \& Test}, vol.~34, no.~4, pp. 7--17, 2017.

\bibitem{kelarev2019multistage}
A.~Kelarev, X.~Yi, S.~Badsha, X.~Yang, L.~Rylands, and J.~Seberry, ``A
  multistage protocol for aggregated queries in distributed cloud databases
  with privacy protection,'' \emph{Future Generation Computer Systems},
  vol.~90, pp. 368--380, 2019.

\bibitem{narayanan2008robust}
A.~Narayanan and V.~Shmatikov, ``Robust de-anonymization of large sparse
  datasets,'' in \emph{2008 IEEE Symposium on Security and Privacy (sp
  2008)}.\hskip 1em plus 0.5em minus 0.4em\relax IEEE, 2008, pp. 111--125.

\bibitem{sweeney2000simple}
L.~Sweeney, ``Simple demographics often identify people uniquely,''
  \emph{Health (San Francisco)}, vol. 671, no. 2000, pp. 1--34, 2000.

\bibitem{ohm2009broken}
P.~Ohm, ``Broken promises of privacy: Responding to the surprising failure of
  anonymization,'' \emph{UCLA l. Rev.}, vol.~57, p. 1701, 2009.

\bibitem{dwork2006calibrating}
C.~Dwork, F.~McSherry, K.~Nissim, and A.~Smith, ``Calibrating noise to
  sensitivity in private data analysis,'' in \emph{Theory of cryptography
  conference}.\hskip 1em plus 0.5em minus 0.4em\relax Springer, 2006, pp.
  265--284.

\bibitem{giraldo2020adversarial}
J.~Giraldo, A.~Cardenas, M.~Kantarcioglu, and J.~Katz, ``Adversarial
  classification under differential privacy,'' in \emph{Network and Distributed
  Systems Security (NDSS) Symposium 2020}, 2020.

\bibitem{lecuyer2019certified}
M.~Lecuyer, V.~Atlidakis, R.~Geambasu, D.~Hsu, and S.~Jana, ``Certified
  robustness to adversarial examples with differential privacy,'' in \emph{2019
  IEEE Symposium on Security and Privacy (SP)}.\hskip 1em plus 0.5em minus
  0.4em\relax IEEE, 2019, pp. 656--672.

\bibitem{cohen2019certified}
J.~Cohen, E.~Rosenfeld, and Z.~Kolter, ``Certified adversarial robustness via
  randomized smoothing,'' in \emph{International Conference on Machine
  Learning}.\hskip 1em plus 0.5em minus 0.4em\relax PMLR, 2019, pp. 1310--1320.

\bibitem{liu2018practical}
Y.~Liu, W.~Guo, C.-I. Fan, L.~Chang, and C.~Cheng, ``A practical
  privacy-preserving data aggregation (3pda) scheme for smart grid,''
  \emph{IEEE Transactions on Industrial Informatics}, vol.~15, no.~3, pp.
  1767--1774, 2018.

\bibitem{ferrag2016survey}
M.~A. Ferrag, L.~A. Maglaras, H.~Janicke, and J.~Jiang, ``A survey on
  privacy-preserving schemes for smart grid communications,'' \emph{arXiv
  preprint arXiv:1611.07722}, 2016.

\bibitem{lu2012eppa}
R.~Lu, X.~Liang, X.~Li, X.~Lin, and X.~Shen, ``Eppa: An efficient and
  privacy-preserving aggregation scheme for secure smart grid communications,''
  \emph{IEEE Transactions on Parallel and Distributed Systems}, vol.~23, no.~9,
  pp. 1621--1631, 2012.

\bibitem{lu2014toward}
R.~Lu, H.~Zhu, X.~Liu, J.~K. Liu, and J.~Shao, ``Toward efficient and
  privacy-preserving computing in big data era,'' \emph{IEEE Network}, vol.~28,
  no.~4, pp. 46--50, 2014.

\bibitem{barbosa2016technique}
P.~Barbosa, A.~Brito, and H.~Almeida, ``A technique to provide differential
  privacy for appliance usage in smart metering,'' \emph{Information Sciences},
  vol. 370, pp. 355--367, 2016.

\bibitem{sandberg2015differentially}
H.~Sandberg, G.~D{\'a}n, and R.~Thobaben, ``Differentially private state
  estimation in distribution networks with smart meters,'' in \emph{2015 54th
  IEEE conference on decision and control (CDC)}.\hskip 1em plus 0.5em minus
  0.4em\relax IEEE, 2015, pp. 4492--4498.

\bibitem{pinte2015low}
B.~Pinte, M.~Quinlan, and K.~Reinhard, ``Low voltage micro-phasor measurement
  unit ($\mu$pmu),'' in \emph{2015 IEEE Power and Energy Conference at Illinois
  (PECI)}.\hskip 1em plus 0.5em minus 0.4em\relax IEEE, 2015, pp. 1--4.

\bibitem{liu2011false}
Y.~Liu, P.~Ning, and M.~K. Reiter, ``False data injection attacks against state
  estimation in electric power grids,'' \emph{ACM Transactions on Information
  and System Security (TISSEC)}, vol.~14, no.~1, pp. 1--33, 2011.

\bibitem{rahman2013false}
M.~A. Rahman and H.~Mohsenian-Rad, ``False data injection attacks against
  nonlinear state estimation in smart power grids,'' in \emph{2013 IEEE Power
  \& Energy Society General Meeting}.\hskip 1em plus 0.5em minus 0.4em\relax
  IEEE, 2013, pp. 1--5.

\bibitem{bhattacharjee2020block}
A.~Bhattacharjee, S.~Badsha, A.~R. Shahid, H.~Livani, and S.~Sengupta,
  ``Block-phasor: A decentralized blockchain framework to enhance security of
  synchrophasor,'' in \emph{2020 IEEE Kansas Power and Energy Conference
  (KPEC)}.\hskip 1em plus 0.5em minus 0.4em\relax IEEE, 2020, pp. 1--6.

\bibitem{bhattacharjee2020blockchain}
A.~Bhattacharjee, S.~Badsha, and S.~Sengupta, ``Blockchain-based secure and
  reliable manufacturing system,'' in \emph{2020 International Conferences on
  Internet of Things (iThings) and IEEE Green Computing and Communications
  (GreenCom) and IEEE Cyber, Physical and Social Computing (CPSCom) and IEEE
  Smart Data (SmartData) and IEEE Congress on Cybermatics (Cybermatics)}.\hskip
  1em plus 0.5em minus 0.4em\relax IEEE, 2020, pp. 228--233.

\bibitem{hossain2020porch}
M.~T. Hossain, S.~Badsha, and H.~Shen, ``Porch: A novel consensus mechanism for
  blockchain-enabled future scada systems in smart grids and industry 4.0,'' in
  \emph{2020 IEEE International IOT, Electronics and Mechatronics Conference
  (IEMTRONICS)}.\hskip 1em plus 0.5em minus 0.4em\relax IEEE, 2020, pp. 1--7.

\bibitem{giraldo2017security_2}
J.~Giraldo, A.~A. Cardenas, and M.~Kantarcioglu, ``Security vs. privacy: How
  integrity attacks can be masked by the noise of differential privacy,'' in
  \emph{2017 American Control Conference (ACC)}.\hskip 1em plus 0.5em minus
  0.4em\relax IEEE, 2017, pp. 1679--1684.

\bibitem{farokhi2018security}
F.~Farokhi and P.~M. Esfahani, ``Security versus privacy,'' in \emph{2018 IEEE
  Conference on Decision and Control (CDC)}.\hskip 1em plus 0.5em minus
  0.4em\relax IEEE, 2018, pp. 7101--7106.

\bibitem{pokhrel2020qos}
S.~R. Pokhrel, Y.~Qu, and L.~Gao, ``Qos-aware personalized privacy with
  multipath tcp for industrial iot: Analysis and design,'' \emph{IEEE Internet
  of Things Journal}, vol.~7, no.~6, pp. 4849--4861, 2020.

\bibitem{jeon2011qos}
Y.-H. Jeon, ``Qos requirements for the smart grid communications system,''
  \emph{International Journal of Computer Science and Network Security},
  vol.~11, no.~3, pp. 86--94, 2011.

\bibitem{pignati2015real}
M.~Pignati, M.~Popovic, S.~Barreto, R.~Cherkaoui, G.~D. Flores, J.-Y.
  Le~Boudec, M.~Mohiuddin, M.~Paolone, P.~Romano, S.~Sarri \emph{et~al.},
  ``Real-time state estimation of the epfl-campus medium-voltage grid by using
  pmus,'' in \emph{2015 IEEE Power \& Energy Society Innovative Smart Grid
  Technologies Conference (ISGT)}.\hskip 1em plus 0.5em minus 0.4em\relax IEEE,
  2015, pp. 1--5.

\bibitem{yang2012man}
Y.~Yang, K.~McLaughlin, T.~Littler, S.~Sezer, E.~G. Im, Z.~Yao, B.~Pranggono,
  and H.~Wang, ``Man-in-the-middle attack test-bed investigating cyber-security
  vulnerabilities in smart grid scada systems,'' 2012.

\bibitem{yu2015towards}
W.~Yu, D.~An, D.~Griffith, Q.~Yang, and G.~Xu, ``Towards statistical modeling
  and machine learning based energy usage forecasting in smart grid,''
  \emph{ACM SIGAPP Applied Computing Review}, vol.~15, no.~1, pp. 6--16, 2015.

\bibitem{du2018differential}
M.~Du, K.~Wang, Z.~Xia, and Y.~Zhang, ``Differential privacy preserving of
  training model in wireless big data with edge computing,'' \emph{IEEE
  transactions on big data}, vol.~6, no.~2, pp. 283--295, 2018.

\bibitem{wang2020edge}
T.~Wang, Y.~Mei, W.~Jia, X.~Zheng, G.~Wang, and M.~Xie, ``Edge-based
  differential privacy computing for sensor--cloud systems,'' \emph{Journal of
  Parallel and Distributed computing}, vol. 136, pp. 75--85, 2020.

\end{thebibliography}

\end{document}